\documentstyle{mn}

\newcommand{\be}{\begin{equation}}
\newcommand{\ee}{\end{equation}}

\newcommand{\Alfven}{ Alfv\'{e}n }
\newcommand{\ba}{\begin{eqnarray}}
\newcommand{\ea}{\end{eqnarray}}
\newcommand{\nn}{\mbox{} \nonumber \\ \mbox{} }
\title{Radio and X-ray Signatures of Merging Neutron Stars}
\author[B. M. S. Hansen \& M. Lyutikov]
{Brad M. S. Hansen\thanks{Current Address: Dept. of Astrophysical Sciences, \newline
 Princeton University;
email:{\bf hansen@astro.princeton.edu}}
 \& Maxim Lyutikov\thanks{email:{\bf lyutikov@cita.utoronto.ca}} \\
 Canadian Institute for Theoretical Astrophysics, University of
Toronto, Toronto, ON M5S 3H8, Canada}
\date{}

\pagerange{\pageref{firstpage}--\pageref{lastpage}}
\pubyear{1999}
\begin{document}

\maketitle
\label{firstpage}

\begin{abstract}
We describe the possible electromagnetic signals expected from the magnetospheric
interactions of a neutron star binary prior to merger. We find that both radio and X-ray
signals of detectable strength are possible. We discuss possible links with the
phenomenon of gamma-ray bursts and describe the prospects for direct detection of these signals
in searches for radio and X-ray transients.
\end{abstract}

\section{Introduction}

The gravitational wave-induced merger of binary neutron stars has evoked considerable interest
in recent years due to their importance as a source of gravitational radiation
 (Thorne 1987 and references therein) and potentially
also gamma-ray bursts (Blinnikov et al 1984; Paczynski 1986; Eichler et al 1989; Paczynski 1991). 
The goal of identifying electromagnetic signatures of the merger events 
is  an important one, even if
such manifestations are not gamma-ray bursts. Given the considerable information processing required
to infer the presence of a gravitational wave burst (Cutler et al 1993), the presence of another
signature will be invaluable.

In this paper we examine the magnetospheric interactions in 
 merging  neutron star binary systems
and describe their pre- and post-merger signatures. In particular, we consider systems containing
one low field $B_r \sim 10^{9-11} G$, rapidly
spinning ($P \sim 1-100 ms$) recycled pulsar and one high field ($B_m \sim 10^{12-15} G$), slowly spinning
($P \sim 10-1000 s$) non-recycled pulsar, as expected on both empirical and evolutionary grounds.
We shall examine how energy is extracted from the spin and orbital motion of the pulsar and in what
manner it is radiated.
Aspects of this calculation have been considered before by Vietri (1996), who considered magnetospherically
induced gamma-ray bursts, and Lipunov \& Panchenko (1996), who
considered the the far-field dipolar and quadrupolar configurations of
a dipole merging with a superconducting sphere.
 Our default estimates will be for systems in which the high field
pulsar has a field $\sim 10^{15}$~G (sometimes called a magnetar),
 which has the potential for the strongest signal. Recent work
suggests that such pulsars may constitute $\sim 10\%$ of the young pulsar population (Kulkarni \& Thompson 1998).

In section~\ref{Extract} we will describe the magnetospheric interactions that remove energy from the orbit
and which allow it to emerge in electromagnetic form. This section will draw heavily on concepts from pulsar
electrodynamics and also the field of satellite-magnetosphere electrodynamics, such as in the Io-Jupiter system.
One of the primary results is that much of the energy is released as  a  pair plasma into the 
 magnetosphere. Section~\ref{Cooling} describes the physical state and evolution of this plasma,
drawing on concepts developed to describe Soft Gamma Repeaters and section~\ref{Discuss} reviews the
state of the observations appropriate to this phenomenon.

\section{The Extraction of Spin and Orbital Energy}
\label{Extract}
High magnetic field pulsars spin down rapidly, so that we consider the high field pulsar to be
essentially non-rotating. A corollary to this is that the light cylinder radius for the 
magnetar magnetosphere (or `magnetarsphere'!) is $\sim c/\Omega \sim 5 \times 10^{11} {\rm cm}
 (P/100{\rm s})$, so that all of our subsequent discussion concerns processes occurring deep within
the closed region of this magnetosphere. This will hold true right up to the point of merger
as tidal interactions cannot enforce synchronisation in a coalescing neutron star binary (Bildsten \&
Cutler 1992). 

The extraction of energy from the pulsar spin and orbital motion is driven by how the strongly
conducting neutron star interacts with the external magnetic field of the magnetar.
 As a 
model problem we consider  perfectly conducting sphere moving through an externally imposed
uniform magnetic field ${\bf B}_0 $ with velocity $ {\bf v }$ and rotating with angular velocity
 $\bf \Omega$.
 Motion of a conducting sphere through
magnetic field is possible only if the resistivity of the sphere
 (neutron star) is nonzero. But the neutron star crust is virtually a perfect conductor:
 the magnetic  diffusion
times for neutron stars are very long - in fact comparable to the age of the universe.
We argue that the required resistivity is due to the dissipation of the
induced magnetospheric currents far from the neutron star surface.
This is analogous to the case of isolated pulsars, where currents are dissipated in the
pulsar wind-ISM shocks more than $10^9$ light cylinder radii away.

The electrodynamics of the low-field pulsar interaction with the magnetar
magnetic field is considered in appendix~\ref{AppA}. Qualitatively, this interaction
has  several important ingredients.  The
conducting neutron star excludes the external field from its interior. The induced magnetic field
has a dipole structure with the magnetic dipole directed against the external field. 
The resultant magnetic field is
\be
{\bf B}_{tot} = {\bf B}_0 + \frac{R^3}{2 r^3}{\bf B}_0 - \frac{3 R^3 (\bf{B}_0 \cdot{\bf r}) {\bf r}}{2 r^5}
\label{Btot}
\ee
At the surface
the total magnetic field has only a tangential component, inside the star the magnetic field
is zero.
The orbital motion of the neutron star with respect to the external field
 will induce surface charges  with a dipole structure 
and surface  charge density  
\be
\sigma _{orb} = { 1 \over 4 \pi c R_0} \left( {\bf B}_0 \cdot [ {\bf R}_0 \times {\bf v} ] \right)
\ee
where $R_0$ is the radius of the neutron star.
These surface charges will also produce electric fields which will have a component along the
total magnetic field. If the neutron star orbits in a vacuum,
 this electric field will accelerate charges from the surface or surrounding region to relativistic energies. 
If the internal magnetic field of the neutron star is exactly zero, the resulting 
structure of the electric field will be of the "outer gap type" (Chen, Ho \& Ruderman 1986) - a region 
in the magnetosphere with $E_{\parallel}$ which does not intersect the surface of the neutron star.
 However, there is likely to be some small component of radial
magnetic field at the surface. This could result either from whatever intrinsic field the
recycled neutron star possesses or from a second-order induced field resulting 
from rotation of the star (see below and appendix \ref{AppA}). 
In this case the electric field will draw charges from the surface.

Similarly to orbital motion,  the rotation of the neutron star will produce a surface charge density
\be
\sigma_{rot}  = \frac{3 R_0 \Omega B_0 \sin \psi \sin \theta}{8 \pi c}
\ee
where $\sin \psi = \cos \Omega t \cos \theta \sin \alpha - \sin \theta \cos \alpha$ and
${\bf \Omega}=\Omega ( \sin \alpha, 0, \cos \alpha)$, i.e. $\psi$ is the polar angle
in the frame aligned with $\bf \Omega$  and rotating with the neutron star. 
This charge density is stationary in the frame of the
neutron star while  in the  laboratory frame  it  yields an additional surface
current ${ \bf j = \sigma}_{rot} { \bf\Omega} \times {\bf r}$. The magnetic field due to this current
is of order $ (R \Omega/c)^2$ smaller than the external field ${\bf B}_0$, but  has a radial 
component at the surface.

Similarly to the case  of the aligned rotator
studied by Goldreich \& Julian (1969), the 
strong  electric field produced by surface charges will accelerate charges 
 in an attempt
to short out the component of the electric field that lies parallel to the magnetic field.
The typical densities of the primary beam will be $n_{GJ} \sim  \Omega  B_0 / 2 \pi e c$
for acceleration by $\sigma_{rot}$ and $ \sim v B_0 / e c R$ for
acceleration by $\sigma _{orb}$.
After being accelerated to sufficient energies ($ \gamma \sim 10^6 $) the
 initial primaries produce curvature photons and a dense  population of secondary 
electron-positron  pairs that will screen the 
induced electric field. 
This 
 mechanism of energy extraction is essentially the same as in the classical pulsar case with
 a couple of small but important differences. The first is that, unlike the case of the pulsar, the 
 near field energy density is dominated by the plasma, rather than Poynting flux (see appendix~\ref{AppA}).
 Secondly, the field configuration defined
by (\ref{Btot}) contains no closed magnetosphere. In the traditional pulsar case, the `working surface' of
the energy extraction is limited to the polar cap, a fraction $\sim (r \Omega/c)^2$ of the the total surface
area, which is linked to the open field lines. In the case under discussion here, the polar cap 
effectively encompasses the
entire star.

The 
energy extracted by accelerating primary particles is limited by the maximum energy that
primary particles can reach:
\be
L \sim 4 \pi R^2 n_{GJ} \gamma_{max} m_e c^3 \sim 3.1 \times 10^{36} {\rm ergs \, s^{-1}}.
\label{Lbeam}
\ee
However, the energy extraction from the orbital motion is likely to be significantly more
efficient than implied by (\ref{Lbeam}). Once the pair production cascade has loaded
 the
external field lines with  plasma, the spiraling neutron star emits \Alfven waves along the external
magnetic field (Drell, Foley \& Ruderman 1965;  Barnett \& Olbert 1986;
Wright \& Southwood 1987), in much same way as Io interacting with Jupiter or various artificial satellites
in the earth's magnetosphere. In this case, the pair production front acts as a surface of finite resistivity,
allowing the neutron star to `cross field lines'\footnote{Even if the resistivity were considerably lower, a 
similar level of energy extraction would occur via the screw-instability of strongly wound magnetic field
configurations (Low 1986; Aly 1991; Volwer, van Oss \& Kuijper 1993; Gruzinov 1999), given only the assumption
of sufficient ambient plasma to justify the force-free approximation.}
 We assume that these waves are dissipated in the magnetar magnetosphere
by non-linear damping mechanisms similar to those invoked by  Soft Gamma Repeater models (e.g. Thompson \& Duncan
 1995).
Thus we shall assume that the bulk of the energy extracted from the orbit is deposited into the magnetospheric
pair plasma, and is of order (Drell et al 1965)
\ba
L_{orb}  \sim&& 4  \pi R^2 B_m^2 \left(\frac{R}{a}\right)^6 \frac{v^2}{c} \nonumber \\
 \sim && 7.4 \times 10^{45} {\rm ergs \, s^{-1}} \left(\frac{B_m}{10^{15} {\rm G} }\right)^2
 \left(\frac{a}{10^{7} {\rm cm}}\right)^{-7}. \label{Lorb}
\ea
Additional sources  of energy are the Poynting losses due to the  motion of the induced  dipole 
(Lipunov \& Panchenko 1996)
and the time varying component of the induced magnetic fields (see appendix \ref{AppA}), 
though the corresponding luminosities
are much smaller than that given by Eq. (\ref{Lorb}).
 Poynting fluxes will be in a form of low frequency electromagnetic waves.
Unlike the equivalent situation for pulsars,  where the density of the secondary plasma is low,
these low frequency electromagnetic waves may not be able to propagate through the dense secondary
plasma present  in the "magnetarsphere" - they will  convert their
energy into plasma
(Asseo et al. 1978).
Thus, most of the energy lost by the neutron star will be converted into plasma in the near zone
 and later
radiated - this is in contrast to normal pulsars where most of the losses within the light
cylinder are due to the
Poynting flux.

Our situation also differs somewhat from that considered by
Vietri (1996), who addressed the problem of the merger of two high-field pulsars.
The consequently large radii of field line curvature  implied
 screening was ineffective and allowed efficient acceleration
of particles and high energy emission.
The fundamental difference in our case is that plasma screening occurs close to the low
field neutron star, where the radius of curvature is smaller. The result is efficient generation
of pair plasma and effective screening of parallel electric fields.
The pair plasma will then mediate the dominant energy extraction by \Alfven wave emission. 

\subsection{Radio Emission}

In normal pulsars, the acceleration of particles by electric fields at the surface
 yields coherent radiation observed
as radio emission. Thus, we might hope for similar signals in this instance. The lack of
a complete theory of pulsar radio emission forces us to adopt a simple parameterisation
based on what we know about pulsars.
 We expect the radio emission to be associated  with the primary beam only, whose
luminosity is given by Eq. (\ref{Lbeam}).
We shall adopt an efficiency
of $\epsilon \sim 0.1$ for the conversion of primary beam energy to radio luminosity, based
on the radio efficiencies of pulsars (see Taylor, Manchester \& Lyne 1993), assuming the pulsar
beam luminosity is
$\sim 10^{-3}$ of the spin-down luminosity (Kennel \& Coroniti 1984).
Then  
 an optimistic estimate for the radio flux at 400~MHz (chosen because it is at this frequency
that the pulsar fluxes are best estimated) is
\be
F_{\nu} \sim 2.1 \, {\rm mJy} \, \frac{\epsilon}{0.1} \left(\frac{D}{100 Mpc}\right)^{-2} B_{15}^{2/3} a_7^{-5/2}. \label{Radio}
\ee
This is within the range of the larger radio telescopes operating today, although somewhat less than the sensitivities
of current radio transient searches.

There are several complications that may preclude generation  of radio emission.
If the neutron star is moving through
a pre-existing plasma generated by the previous orbital cycles
the electric gaps may be quenched, 
 there will be no
need to accelerate further particles and the beam luminosity may drop to zero.
In addition, the formation of positronium in the magnetic fields  exceeding
 $\sim 4 \times 10^{12}~G$ (Usov \& Melrose 1996, Arons 1998) may also  quench  the radio emission.

Furthermore, the  generated radio emission may be absorbed in the  magnetarsphere. 
 We expect that {\em nonresonant} Thomson
scattering of the low frequency ($\nu << \nu_B$) radio emission will not be important
due to the the strong  suppression ($\sigma = \sigma_T ( \nu/\nu_B)^2$)  of the scattering
cross-section by the magnetic field at low frequencies. Resonant cyclotron absorption may be
important in the outer regions of the magnetarsphere where the cyclotron frequency becomes
comparable to the radio wave frequency ($a \sim 10^{10} cm$). Nevertheless, such absorption does
not occur in the pulsar case, so we may reasonably expect the radio emission to escape the 
magnetarsphere.

Thus, the first electromagnetic signature we anticipate from a realistic merging neutron star
binary is a coherent radio burst, emitted $\sim$~seconds before the gravitational wave burst.
However, the effects of interstellar  dispersion can cause  delays of
hours, possibly allowing for radio follow-up observations at low frequencies
(Palmer 1993; Lipunova, Panchenko \& Lipunov 1997).

\section{Evolution of the Magnetospheric Pair Plasma }
\label{Cooling}

Most of the energy liberated by the strong electric fields of section~\ref{Extract} is not
radiated, but is rather released into the magnetosphere of the slowly-rotating magnetar
in the form of \Alfven waves and a dense pair plasma. The energy release (\ref{Lorb}) is a significant
fraction of
 the local magnetic energy density. In such a case,  a wind, driven either by hydromagnetic or plasma
pressure, is likely to result
 (Paczynski 1986, 1990, Melia \& Fatuzzo 1995; Katz 1996) while some will remain trapped, in a fashion similar to that of the
 Soft Gamma Repeater picture of a magnetically
confined pair plasma (Thompson \& Duncan 1995). We envisage that the plasma released into
regions of decreasing field strength powers the wind while plasma released into regions of increasing
field strength will be trapped. Figure~\ref{Drawing} shows a schematic version of our scenario.

Let us consider first the case of the wind. A release of energy at the rate given by equation (\ref{Lorb})
results in a compactness parameter $\eta = L/a c \sim 10^7 B_{15}^2 a_{-7}$. Thus, this is the same situation
envisaged in cosmological models for gamma-ray bursts (Paczynski 1986; Goodman 1986), wherein the release of
a large quantity of pure energy within a small volume leads to a relativistically expanding fireball.
The energy release during the neutron star inspiral will drive a relativistically expanding wind of pairs
and photons.
 Thermal
and statistical equilibrium between photons and pair plasma is maintained during the
expansion by pair production and comptonization
(Cavallo \& Rees 1978) until the 
 comoving temperature drops to $T \sim 3 \times 10^8$~K and pair production can no longer
maintain the necessary electron scattering optical
depth. At this point the radiation escapes, with an approximately thermal spectrum.
 However, the relativistic boost increases the observed temperature by a factor $\gamma$, the original
Lorentz factor of the fireball and reduces the observed burst time by the same factor. Thus, the observed
energetic and temporal properties of the wind emission may be approximately described by thermal emission
at the appropriate initial temperatures and timescales, despite the fact that the true photosphere is on
scales much larger than the original volume.
 Hence, we shall estimate the observed flux in this case as
\be
F_{wind} \sim \frac{\alpha L}{4 \pi a^2} \sim 3 \times 10^{30} \alpha {\rm ergs \, cm^{-2} \, s^{-1}} B_{15}^2 a_7^{-9},
\label{Fwind}
\ee
($\alpha$ is the fraction of the energy release lost in the wind) yielding effective temperatures just before merger $\sim 1.5 {\rm MeV} B_{15}^{1/2}$.

The case of the trapped plasma is somewhat more subtle. This plasma is 
 very optically thick (Svensson 1987; Thompson \& Duncan 1995)
and the 
 inspiral time $\sim 0.4 \, {\rm s} \, a_7^4$ is short. Hence, very little of the total energy contained
in the magnetosphere is radiated  in this time. The emission that does occur is dominated by the region
just above the surface of the magnetar, where the strong magnetic field decreases the electron scattering
cross-section and thereby promotes a larger photon flux. At late times, the plasma temperature is high
enough that ablation of material from the surface of the star is likely to provide an Eddington limit
\be
F_{edd} \sim 4.14 \times 10^{26} {\rm ergs \, cm^{-2} s^{-1}} B_{15} a_7^3. \label{Fedd}
\ee
In the case of Eddington limited cooling, we see that the emission actually gets softer as
the inspiral proceeds (the opposite of the contribution (\ref{Fwind}) from the relativistic
wind) because the plasma temperature increases and acts to negate the magnetic suppression
of the electron scattering cross-section.

Thus, our strongest prediction is the presence of an X-ray precursor to the neutron star
merger. This precursor should be dominated by approximately thermal emission from the wind component, which
progressively hardens as the binary approaches merger. For magnetars ($B_{15} \sim 1$), this
can approach  energies $\sim$~1.5~MeV while binaries containing  a normal pulsar 
($B_{15} \sim 10^{-3}$) will only get as hard as $\sim 50$~keV. In some cases these objects
may be accompanied by softer ($\sim$~1~keV) components associated with the cooling of
the trapped plasma.
The  observed flux will be dominated by the harder wind component, with flux levels
$\sim 3 \times 10^{-9} {\rm ergs \, cm^{-2} s^{-1}} B_{15}^2 a_7^{-7}$ for a source
at 100~Mpc (the distance scale for which we expect a few neutron star mergers per year).

\subsection{The Ultimate Fate of the Pair Plasma}

Much of the energy released by the processes in section~\ref{Extract}
is retained in the magnetospheric plasma for timescales longer than the inspiral time, i.e.
the merger event will occur surrounded by a significant magnetospheric plasma.
 This energy totals about
$E_{\rm plasma} \sim 2 \times 10^{47} {\rm ergs} B_{15}^2$. Vietri (1996) proposed that this
energy, most of which is released on the last few orbits, could power a Gamma-Ray Burst as it
escapes it's magnetic confinement.
Our estimate of the energy release is somewhat smaller than his (strictly speaking he calculated
the maximum energy a magnetosphere could contain, rather than the energy release itself).
Recent determinations of the distances (Metzger et al 1997; Kulkarni et al 1998; Djorgovski et al 1998)
 to burst events suggest that much larger energy
releases are required to explain many gamma-ray bursts (modulo beaming considerations). 

A more intriguing possibility occurs if it is the magnetar which is disrupted to form a rapidly
rotating disk around the compact merger remnant (as might be expected from the non-recycled and
thus presumably lighter object). If the magnetic field footpoints remain tied to
the disrupted material the magnetosphere is forced to co-rotate with the 
disk and the corotation radius must move rapidly inwards, converting closed field lines to
open ones and ejecting the magnetospheric plasma. This would allow the plasma to tap the much
larger reservoir of disk rotational energy to power the gamma-ray burst (as in many other
gamma ray burst models. See Hartmann 1996 for a review) and could also account
for significant collimation of the outflow.
Furthermore, the approximate equipartition between plasma energy
density and magnetic field energy is appropriate for the formation of an episodic jet (Ouyed \& Pudritz 1997),
which may contribute to burst temporal variability.
Finally, the magnetospheric origin of the pair plasma would also avoid
the baryonic loading problem encountered by mechanisms which propose to generate the pairs
by neutrino annihilation close to the merger product (Janka \& Ruffert 1996; Ruffert \& Janka 1998).

\section{Discussion}
\label{Discuss}
\subsection{Observations}
Our results predict the appearance of X-ray and Radio transients as precursors to gravitational
wave bursts and possibly also Gamma-Ray Bursts.

Motivated by the search for X-ray counterparts to GRB, Gotthelf, Hamilton \& Helfand (1996)
and Greiner (1999) have searched the Einstein \& ROSAT databases, respectively, for brief
X-ray transients. The energy ranges searched are somewhat softer ($\sim 0.1-4$~keV) than we would predict for the
peak of the energy distribution. Nevertheless, both searches found classes of transients 
 of possible astronomical origin in sufficient abundance to encompass any reasonable estimate
of the event rate (Phinney 1991; Narayan, Piran \& Shemi 1991; Lipunov et al 1995,
Arzoumanian, Cordes \& Wasserman 1999). Searches for untriggered bursts in the BATSE catalogue 
(Kommers et al 1997) proved even more interesting, revealing a class of bursts restricted to
the 25-50~keV channel. Most of these events are consistent with being intensity fluctuations in
Cygnus~X-1, but the remaining 10\% are consistent with the expectations of our model.

If some neutron star mergers do yield GRB, then our results may provide an explanation for the
subset of GRB discovered to show X-ray precursors (Murakami et al 1992; Castro-Tirado et al 1993; in 'T Zand
et al 1999). These events show the characteristic soft-to-hard spectral evolution we anticipate, although
the spectrum in well-studied cases such as GB900126 (Murakami et al 1991), appears somewhat softer ($\sim 1.5$~keV)
than we would expect. The flux ($\sim 2 \times 10^{-9} {\rm ergs \, cm^{-2} s^{-1}}$), however, is 
appropriate, suggesting perhaps that an analysis more sophisticated than the black-body assumption
 is required.

Perhaps the best candidate for our model is the unusual transient GB900129 observed by Ginga (Strohmayer et al 1995),
which yielded a thermal bremstrahlung temperature $\sim 20$~keV and duration 5-10~s. Strohmayer et al note
the similarity to the SGR spectral characteristics, which agrees with the magnetospheric origins in our
model as well. Figure~\ref{Xray} shows the comparison of the energetics of various observed transients with our models.

Radio transients associated with GRB are a subject of growing interest and several searches (e.g. FLIRT and STARE)
are ongoing. However, radio searches for brief transients are particularly bedevilled by terrestrial
interference. Most limits lie in the 10-100~kJy range at 76~MHz (FLIRT; Balsano 1999) and 611~MHz (STARE; 
Katz et al 1998). These don't particularly constrain our model, which anticipates signals $\sim$~mJy-Jy. One
radio transient uncovered by FLIRT does deserve mention. FLIRT (Balsano 1999) located a radio transient 
apparantly associated with GRB~980329. The transient is unique in the database and occurred within 50~s of
the burst. The transient showed evidence for dispersion, with a DM$\sim 66 {\rm pc \, cm^{-3}}$. All these
argue that the association is real. However, the signal was very narrow band, indicative of terrestrial
interference. If one does choose to interpret this ambiguous transient as a real association, the dispersion
measure would rule out a truly cosmological burst. Furthermore, the $\sim$~kJy flux would suggest
distances $\sim$1~Mpc based on our luminosity estimates. All these would argue against the suggestion that
the event occurred at very high redshift (Fruchter 1999) and the red optical transient would most likely
arise from extinction (Reichart et al 1999).

\subsection{Binaries with Black Holes}

We have not yet discussed the signatures of mergers associated with binaries in which 
one of the components is a black hole rather than a neutron star, although such binaries
may outnumber the double neutron star binaries (e.g. Bethe \& Brown 1998). If the black hole
is formed from a strongly magnetized object, then only open field lines remain. Thus, the
inspiral of a neutron star through this magnetosphere will still generate the relativistic
wind and it's associated X-ray signature\footnote{This precursor may also be cut off before the
merger if the open field lines are restricted to only a small polar cap, i.e. the
neutron star may eventually orbit in a field free zone.},
 but there will be no trapped magnetospheric plasma.
We would then expect to see the same X-ray and radio precursor to the event, but no soft X-ray
component to the precursor or any post merger signature associated with the magnetospheric
pair plasma. For binaries in which the low field component is a  black hole,  the effective resistivity
of the event horizon (Thorne, Price \& Macdonald 1986) is considerably larger than that for a neutron star crust. Consequently the 
distortion of the magnetic field due to the orbital motion is much smaller and the energy extraction
in observable energy is similarly reduced. Furthermore, one cannot extract charged particles from the event
horizon, although `outer gap' acceleration is still feasible. We expect such mergers to be (electromagnetically)
much
quieter.

\section{Conclusion}

If neutron star mergers are not associated with GRB, then any additional electromagnetic signatures
will be invaluable when the search begins in earnest for the gravitational wave signal. Li \& Paczynski (1999)
have suggested one such signature; namely a post-merger mini-supernova powered by radioactive decay of disrupted
neutron star material. We have demonstrated the possibility of additional {\em precursor} signals in the
radio and X-ray regimes, driven by the magnetospheric interactions of the neutron star and their magnetic fields.
Our results differ somewhat from those of Vietri (1996) who considered a related model. We ascribe this
to the much more localized interaction in our scenario, the result of a more realistic choice of parameters,
and to our more complete description of the electrodynamics of the accelerated plasma.

To conclude we re-iterate the properties of what we would consider a prime candidate for an electromagnetic
counterpart to a neutron star merger. Estimates of the merger rate suggest that the events typically
observed would be at distances $\sim 100$~Mpc, suggesting X-ray fluxes $\sim 3 \times 10^{-9} {\rm ergs \, cm^{-2} s^{-1}}$
with effective temperatures progressing upwards through the 10-100~keV range preceding the gamma-ray event on timescales
of order seconds or less. Associated radio fluxes could be as much as $\sim 5$~Jy at this distance, although the
 ability of the radio waves to propagate in the late-time plasma shroud is 
rather uncertain. The coincidence of the radio signal could be influenced by dispersion in both the host galaxy and ours.
Dispersion in the inter-Galactic medium will be of the order of $\sim 1 {\rm cm^{-2} pc} (D/100 {\rm Mpc})$ for an ionized IGM mass fraction
$\sim 10^{-2}$ of the critical density and thus is unlikely to contribute significantly for any detectable events.
 There are also several possible signatures of the
merger event itself, depending on how the orbital and binding energies of the binary and components is disbursed between
 remnant and ejecta.

We thank Steve Thorsett, Vicky Kaspii \& Jackie Hewitt for information regarding the FLIRT and STARE radio transient programs
and Vladimir Lipunov and Andrei Gruzinov for discussions.

\clearpage

\begin{appendix}

\section{Electromagnetic interaction of a neutron star with external magnetic field}
\label{AppA}
Our problem concerns the electrodynamics of a low field neutron star orbiting in the field of a
high field neutron star. Although the intrinsic field of the neutron star will be important, let us
begin with the model problem of a spinning, conducting sphere
in a uniform, externally imposed field.

Consider an unmagnetized conducting sphere in an external homogeneous  magnetic field
${\bf B_0} $. 
The external field will induce 
surface currents
\begin{equation}
{\bf g}  = { c \over 4 \pi } { B_0} \sin \theta {\bf e}_ {\phi}
\end{equation}
flowing in the azimuthal direction ${\bf e}_ {\phi}$ about the ${\bf B_0}$ axis. These currents will
 induce magnetic field  with a dipole structure $  {\bf \mu} = - {\bf B_0} R_0^3 /2 $
so that the total field is
\be
{\bf B}_{tot}=
{\bf B_0} + {{{{R_0}^3}\,{\bf B_0}}\over {2\,{r^3}}} - 
  {{3\,{{R_0}^3}\, ({\bf B_0}\cdot{\bf r})\,{\bf r}}\over {2\,{r^5}}}
\ee
At the surface the radial component of the total magnetic field is zero.
Consider now the same sphere, moving with velocity $\bf v$ and rotating with angular
velocity $\Omega$. 
 Up to the relativistic correction of the order   $v^2/c^2$ the magnetic field 
 is the same  in the star's rest frame (which is moving and rotating with respect 
to the laboratory frame).
In the  star's rest frame the electric field is a sum of electric fields due to  the uniform
motion and rotation. The electric field due to the rectilinear motion 
is uniform, given by ${\bf E}_{orb} = {1\over c} \left[  {\bf v} \times  {\bf B}_{0} \right]$, while
the electric field due to the  rotation is 
 ${\bf E}_{rot} ={1\over c}\left[\left[  {\bf \Omega} \times {\bf r}  \right] \times  {\bf B}_{tot} \right]$.
${\bf E}_{rot}$ 
has radial and meridional components:
\begin{eqnarray*}
E_{ rot,r} = 
\frac{\left( 2\,{r^3} + {{R_0}^3} \right)}{2\,c\,{r^4}} 
     \left( {r^2}\, ({\bf B_{tot}} \cdot {\bf \Omega } ) - 
        ({\bf B_{tot}} \cdot{\bf r})\, ( {\bf \Omega } \cdot {\bf r}) \right) 
\end{eqnarray*}
\be
E_{rot, \psi}= 
\frac{{\left( {r^3} - {{R_0}^3} \right)}}{c\,{{r^5}}}  ({\bf B_{tot}} \cdot {\bf r})\,
     \left( -\left( {r^2}\, {\bf \Omega }  \right)  + 
        ( {\bf \Omega } \cdot {\bf r})\,{\bf r} \right) 
\ee
where $\sin \psi=  \left( \cos \phi \,\cos \theta \,\sin \alpha - \cos \alpha \,\sin \theta 
        \right)$
is the polar angle in the frame alinged with ${\bf \Omega}$
and we assumed that ${ \bf B}$ is antiparallel to $ { \bf z }$
and ${ \bf \Omega} = \Omega \{ \sin\alpha,0, \cos \alpha\}$.
There is  also a non-inertial, spatially distributed 
 charge density associated with this electric field 
(the Goldreich-Julian density)
\be
\rho = {1\over 4 \pi e}  {\rm div} {\bf E} = {  {\bf \Omega \cdot B_{tot} } \over 2 \pi e c}.
\ee

The ample supply of charges inside the star will  short out the total electric field
inside the star by 
producing 
surface charge density $\sigma= {1\over 4  \pi} E_r$:
\ba
\sigma _{orb} &=& { 1 \over 4 \pi c R_0 } \left( {\bf B_0} \cdot [ {\bf R}_0 \times {\bf v}] \right)
\nn 
\sigma_{rot}&=&
{{3\,B_{tot}\,  \Omega  \,R_0\,\sin (\psi )\,\sin \theta }\over {8\,\pi c }}
\ea
The surface charge distribution $\sigma _{orb}$ has dipole structure with
${\bf b} = { R_0^2 \over 8 \pi c } \left[ {\bf v} \times {\bf B_0} \right]$, while
$\sigma _{rot}$ has monopole (total charge 
$Q_{rot}= -\,B _0\,  \Omega  \,{{R_0}^3}\,\cos \alpha /c$)
  and quadrupole contributions.
Both types of 
 surface charges  will produce
electric fields outside of the star with nonzero component 
along the magnetic field line of the order
\ba
E_{\parallel, orb} &\approx& {  {R_0}^3 \over c r^4} 
\left( {\bf B} _0 \cdot \left[  {\bf r} \times {\bf v} \right] \right) \cos \theta
\nn
E_{\parallel, rot} &\approx& -{B_0\,  \Omega  \,{{R_0}^3}\,\cos \alpha  \cos \theta \over  c r^2}
\ea
These electric fields will accelerate the primary charges to relativistic energies.

The surface charge  distribution  $\sigma _{orb}$ is stationary in the moving but non-rotating frame,
while surface charges  $\sigma _{rot}$ are stationary in the neutron star frame (which is
moving and rotating with respect to  the lab frame).
An observer in  the neutron star frame will detect three types
of electric currents: due to rotation of  $\sigma _{orb}$ and
inertial currents due to electric fields of the surface charges
${\bf j} _{in}=\left[ ({\bf E}_{ \sigma _{orb}} + {\bf E} _{ \sigma _{rot}} )
\times {\bf \Omega } \right] /( 4 \pi)$.
Inertial current due to ${\bf E} _{ \sigma _{rot}}$ will 
generate  magnetic field  of the order $ B _{rot} \approx B_0 ( \Omega \,R_0 /c )^2 $
with a component perpendicular to the surface of the millisecond pulsar.
Equivalently, in the laboratory frame the charges  $\sigma _{rot}$ rotating with the star
 will produce  surface currents along the  $\psi $ direction
${\bf g}_{rot} = \sigma _{rot}
\left[{\bf \Omega \times r} \right]$  
 that  will generate  magnetic field $ B _{rot}$. 

The Poynting losses  due to rotating dipole will be proportional to the
time varying component of the induced magnetic field 
$  \propto B_0 ( \Omega \,R_0 /c )^2 \sin
\alpha  \cos (\Omega t)$,
so that the resulting  Poynting flux would be $ P \propto B^2 ( \Omega \,R_0 /c )^4 \sin ^2 \alpha$.
It is suppressed by a small factor $( \Omega \,R_0 /c )^4 \ll 1$ if compared with
the rotating dipole with the strength equal to the external magnetic field.

\end{appendix}

\clearpage

\begin{figure}
\caption{Schematic version of the energy extraction process. The motion of the companion
through the magnetar field induces a plasma flow from the companion into the magnetosphere.
The pressure of this flow will drive a relativistic wind in those regions where the
flow moves into a regime of weaker field, while the plasma remains trapped in the case when it
flows into a stronger field regime. The hot pair plasma will ablate some baryons off the surface
of the neutron star, providing a baryon-loaded sheath which regulates the cooling of the
trapped plasma.\label{Drawing}}
\end{figure}

\begin{figure}
\caption[Tt.ps]{Here we show the expected evolution of the X-ray precursor signal
for binaries containing a recycled pulsar with either a normal field neutron star or a magnetar.
 The solid lines
indicate the wind emission (the solid vertical line at the right indicates the luminosity)
while the dotted lines indicate the cooling emission of the trapped magnetospheric
plasma (also with appropriate luminosity scale). The dashed part of the $10^{15}$~G curve
indicates the region where bound positronium  production may alter the plasma injection
mechanisms, possibly resulting in more nonthermal, high-energy emission.
 The points show the various transients
discussed in the text (solid symbols are associated with GRB, open symbols are not).
The time error bars are determined from the quoted time resolution and rise times
of the signals. We see that the detected temperatures and time delays are broadly consistent
with the expected theoretical values.\label{Xray}}
\end{figure}

 
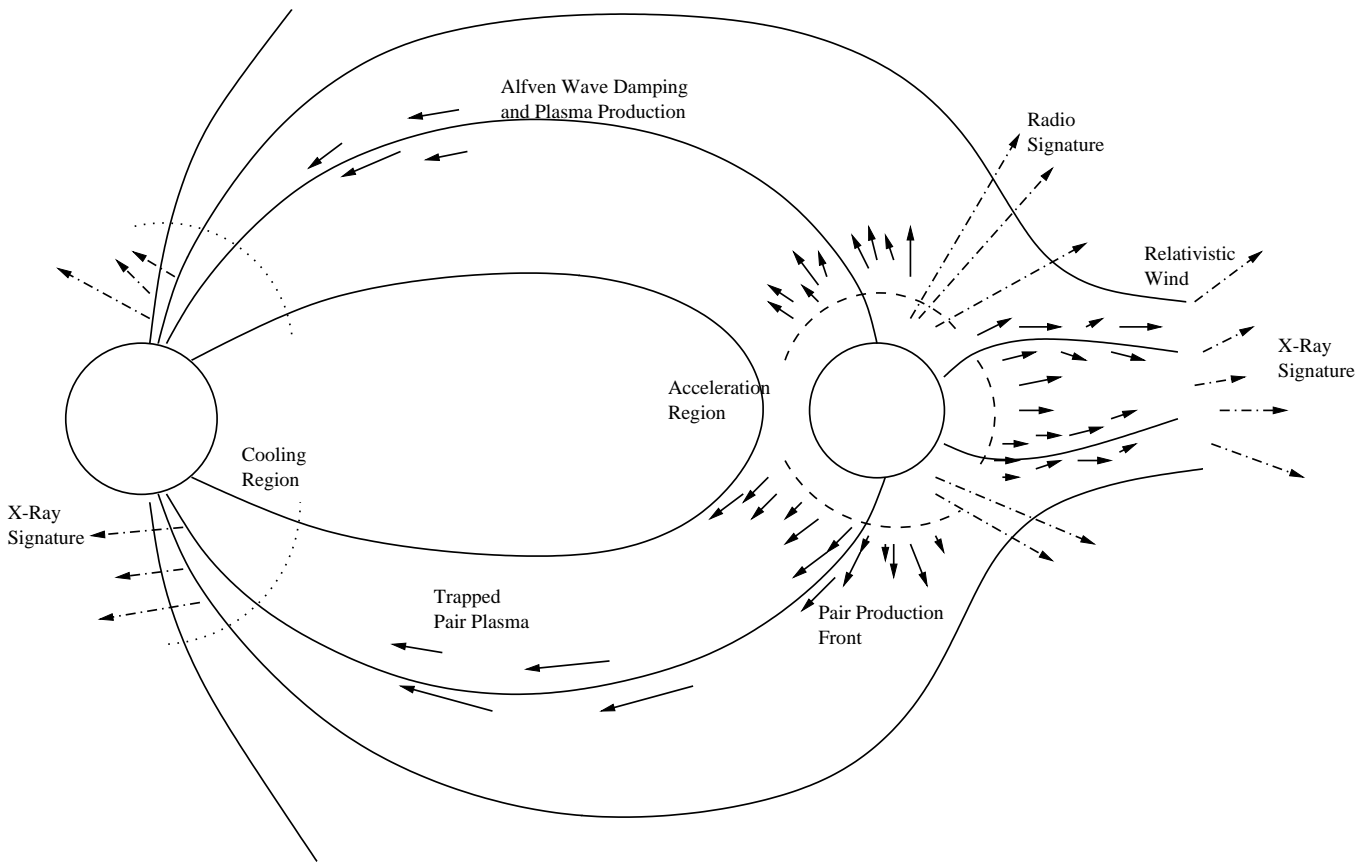
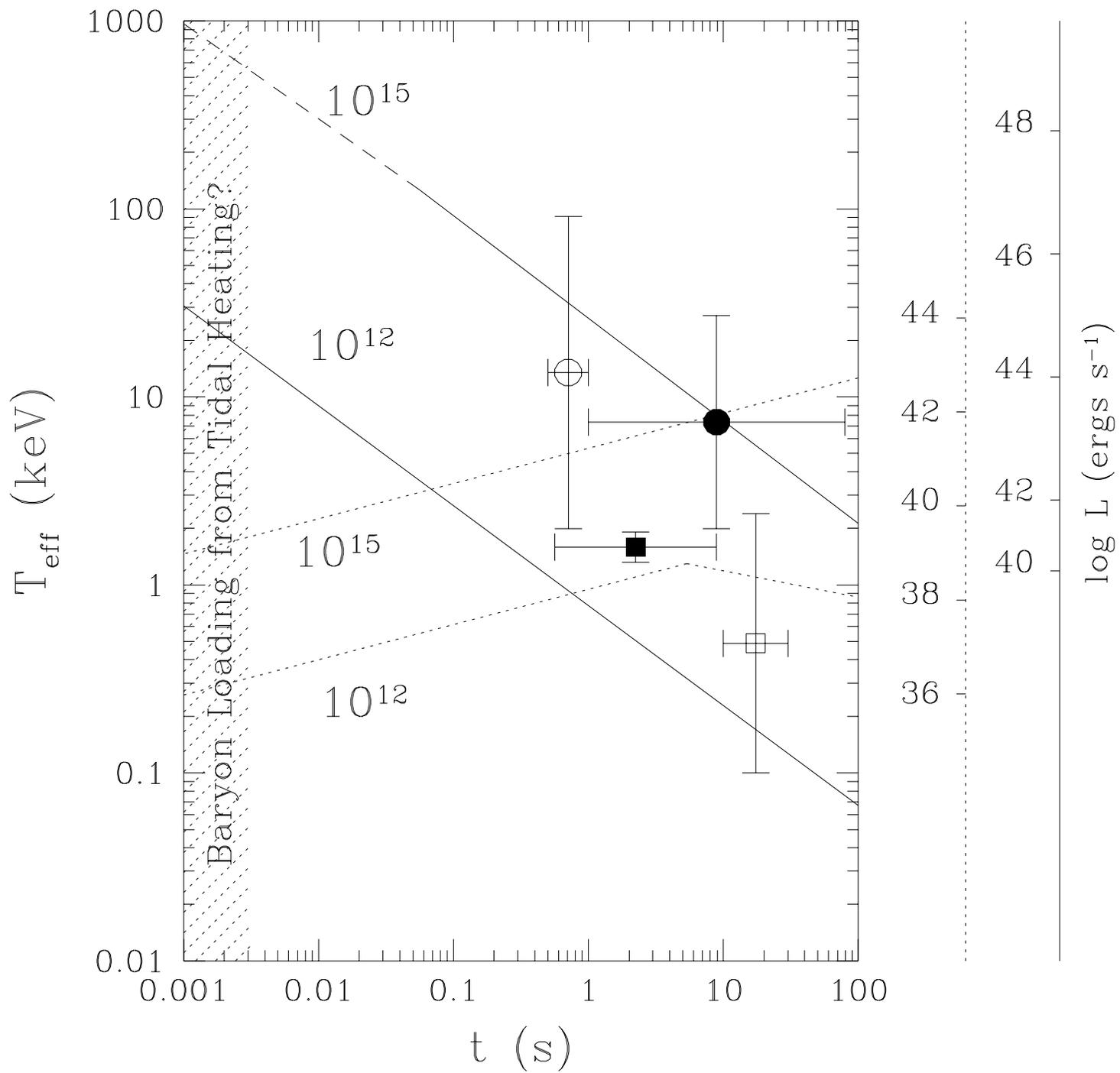
\end{document}